\documentstyle[12pt]{article}
\renewcommand\caption{}

\title{On Color Superconductivity in External Magnetic Field
}

\author{{\sl E. V. Gorbar} \\
{\sl{Instituto de Fisica Teorica, 01405-900 Sao Paulo, Brazil} \dag}
}

\date{}

\begin{document}

\maketitle

\vfill

\begin{abstract} 

We study color superconductivity in external magnetic field.  We discuss the reason
why the mixing angles in color-flavor locked (CFL) and two-flavor
superconductivity (2SC)
phases are different despite the fact that the CFL gap goes to the 2SC gap for $m_s \to \infty$.
Although flavor symmetry is explicitly broken in external magnetic field, we show that
all values of gaps in their coset spaces of possible solutions in the CFL phase
are equivalent in external magnetic field.

\end{abstract}
PACS 12.38.-t, 11.15.Ex, 97.60.Jd \\
\dag On leave of absence from Bogolyubov Institute for Theoretical Physics, 252143,
Kiev, Ukraine

\vspace{2mm}

\vfill
\eject

\newpage

\newpage
\section{Introduction.}

\vspace{1cm}

It was shown some time ago that QCD at high baryon density is a color superconductor \cite{Bar}.
Recently there was a significant activity \cite{activity, strange} in studing color superconductivity
caused mainly by the
observation that the superconducting gap can be as high as 100 MeV.  It is believed that
in nature and laboratory experiments color superconductivity can occur in high energy ion collisions and in the
cores of neutron stars. Since temperatures obtained in high energy ion collisions are usually significantly more
than temperatures characteristic for the color superconductor/quark gluon plasma phase transition at the
relevant baryon densities, it is unlikely that color superconductivity can be observed in high energy ion collisions
\cite{Raja}. Thus, it is possible that perhaps the only chance of observing
color superconductivity is connected with astrophysical observations of neutron stars. Therefore,
the study of astrophysical implications of
the occurence of color superconductivity in the cores of neutron stars is an actual and important problem.

As well known neutron stars typically possess very strong magnetic fields up to $10^{13} G$.  For usual
superconductors, strong enough external magnetic fields destroy superconductivity. Consequently, it is important
to study how the presence of external magnetic field affects the physics of neutron stars
with color superconducting cores. This problem was considered in a recent paper by Alford, Berges, and Rajagopal
\cite{ABR} (for earlier studies and other astrophysical aspects of the occurence of color superconductivity
in the cores of neutron stars see \cite{Bla}). Since there is a 'modified' photon, which remains massless
in color superconducting phase (it is in the general case a mixture of
electromagnetic and some gluon fields), 
Alford, Berges, and Rajagopal showed that external magnetic field
partially penetrates inside color superconducting cores and it is very unlike that magnetic fields typical for
neutron stars can destroy color superconductivity in the cores of neutron stars.

In this paper we continue the study of color superconductivity in external magnetic field.
In Sect.2 we discuss the reason why the mixing angles in CFL and
2SC phases are different despite the fact that the CFL gap goes to the 2SC gap for $m_s \to \infty$.
We show in Sect.3 that although flavor symmetry is explicitly broken in external magnetic field
all values of gaps in their coset spaces of possible solutions in the CFL phase
are equivalent in external magnetic field. Our conclusions are given in Sect.4.

\vspace{1cm}

\section{ Why are mixing angles in the 2SC and CFL phases different? }

\vspace{1cm}

In paper \cite{ABR} the CFL phase with three massless quarks and the 2SC phase with two massless
quarks were explicitly considered. On the other hand it is known that in
the real QCD the mass of the strange quark cannot be neglected \cite{strange}. According to \cite{strange}, it is
expected that for realistic values of $m_s$ a CFL phase with 5 independent order parameters is realized.
Therefore, it is natural to consider explicitly physical properties of the CFL phase with $m_s \ne 0$ in external
magnetic field. (In what follows we consider only the case of the so called 'sharp boundary'
transition between nuclear and superconducting quark matter. The case of 'smooth boundary' transition is
trivial because there is not even partial exclusion of magnetic field flux inside the superconducting region \cite{ABR}.)

According to \cite{strange}, the same linear combination of the electromagnetic and gluon fields, which is
massless in the CFL phase with $m_s=0$, is also massless in the CFL phase with $m_s \ne 0$.
Therefore, the mixing angle of the usual photon with gluon fields in the CFL phase with $m_s \ne 0$
coincides with the mixing angle in the CFL phase with three massless quarks and, consequently, the same part of
magnetic field flux penetrates inside a color superconductor in the CFL phase with $m_s \ne 0$.
However, one question remains to be answered. As shown in \cite{strange}, for $m_s \to \infty$,
the gap in the CFL phase with $m_s \ne 0$ goes to the 2SC gap.
Obviously, extremely heavy strange quark decouples from the low energy
dynamics in this case (of course, in the opposite limit $m_s \to 0$, the gap in the CFL phase with $m_s \ne 0$ goes
to the gap of the CFL phase with three massless quarks). Therefore,
one would naively expect that mixing angle is the same in all three phases. However, according to \cite{ABR},
the mixing angle in the 2SC phase is two times less than in the CFL phase.
Why are mixing angles in these phases different?

To answer this question we first consider in more detail the definition of mixing angle.
The color superconducting gap is a vacuum expectation value of two quark
fields $<\psi^{i}_{\alpha} \gamma_5 C \psi^{i}_{\beta}>$, therefore,
it transforms under flavor and color transformations as a tensor product of two natural representations of
flavor and color groups of symmetry.
If we group indices i and $\alpha$ as well as j and $\beta$, then the gap $\Delta_{\alpha \beta}^{ij}$
is a matrix with respect to two
united indices i,$\alpha$ and j,$\beta$. The action of the covariant derivative on the gap is
\begin{equation}
D_{\mu} \Delta = (\partial_{\mu} + ieA_{\mu}Q_{t.r.} + igA_{\mu}^{a}T^{a}_{t.r.}) \Delta,
\end{equation}
where $t.r.$ means tensor representation, $(Q_{t.r.} \Delta)_{\alpha \beta}^{ij} =
Q^{ii_1}\Delta_{\alpha \beta}^{i_1j} + \Delta_{\alpha \beta}^{ii_1}
(Q^T)^{i_1j}$, \\ $(T^{a}_{t.r.} \Delta)_{\alpha \beta}^{ij} = T^{a}_{\alpha \alpha_1}\Delta_{\alpha_1 \beta}^{ij} +
\Delta_{\alpha \alpha_1}^{ij}(T^{a})^T_{\alpha_1 \beta}$ ($A^T$ means a transpose to a matrix A), Q=diag(2/3, -1/3, -1/3)
the generator of electromagnetic
transformations, $T^{a} = \frac{\lambda^{a}}{2}$ the generators of color transformations, and $\lambda^a$
the Gell-Mann matrices. In the general case to find an operator, which is equal to zero acting on the gap,
we consider an operator of the general form $\tilde{Q} = Q + a_1T^1 +
... +  a_8T^8$ and seek a solution of the equation $\tilde{Q} \Delta = 0$ that gives us the sought operator $\tilde{Q}$. 
By representing $ieA_{\mu}Q + igA_{\mu}^{a}T^{a}$ as
\[
i\left(
\begin{array}{rrrr}
A_{\mu} & A^1_{\mu} & \ldots & A^8_{\mu}
\end{array} \right)
\left(
\begin{array}{r}
eQ \\
gT^1 \\
\vdots \\
gT^8
\end{array} \right)
\]
and inserting $O^TO$, where
\[
\begin{array}{rrr}
 & & \\
 &O&= \\
 & & \\
 \end{array}
\left(
\begin{array}{rrrr}
n_0 & n_1 & \ldots & n_8 \\
\vdots & & & \\
m_0 & m_1 & \ldots & m_8 \\
\end{array} \right)
\]
is an orthogonal 9$\times$9 matrix, we find elements $n_0, n_1,...,n_8$ from the equation
$en_0Q + gn_1T^1 + ... + gn_8T^8 = a\tilde{Q} = a(Q + a_1T^1 + ... +  a_8T^8)$
\begin{equation}
n_0 = a/e, \,\, n_1 = \frac{aa_1}{g} \,\,..., \,\, n_8 = \frac{aa_8}{g}
\end{equation}
and, consequently, the corresponding massless linear combination of electromagnetic and gluon fields is
$\tilde{A_{\mu}} = n_0A_{\mu} + n_1A^1_{\mu} + ... + n_8A^8_{\mu}$.
(It is an easy task to check that $a_3=-1$ and $a_8=-\frac{1}{\sqrt{3}}$ (others $a_i=0$)
is the sought solution in the case of the CFL phase, i.e.,
the operator $\tilde{Q} = Q - (T^3 + \frac{T^8}{\sqrt{3}})$
is equal to zero acting on the CFL gap.) A generalized mixing angle is defined
as $\arccos$ of the element $O_{11}$ of the matrix O, i.e., it is $\alpha=\arccos n_0$. (We say a generalized
mixing angle because a
9$\times$9 orthogonal matrix cannot be parametrized by one independent parameter unlike the familiar
case of mixing of two gauge fields in Standard Model. However, since only the element $O_{11}$ of the
matrix $O$ is important for us, it is convenient to define a generalized mixing angle as $\arccos$ of the
element $O_{11}$).
Since $OO^T=1$ for orthogonal matrices, we have $n_0^2 + n_1^2 + ... + n_8^2 = 1$ that gives us
$a=\frac{eg}{\sqrt{g^2 + (a_1^2 + ... + a_8^2)e^2}}$.
Therefore, for the CFL phase, we find that the mixing angle is $\alpha_{CFL} = \arccos \frac{g}
{\sqrt{g^2 + 4e^2/3}} \approx \frac{2e}{\sqrt{3}g} \approx 1/10$, where we assumed that $a_s = \frac{g^2}{4\pi} \approx 1$
at the scale of baryon densities typical for neutron stars cores and, consequently,
the gauge field $\tilde{A_{\mu}} = \frac{gA_{\mu} -
eA^{3}_{\mu} - \frac{e}{\sqrt{3}}A^{8}_{\mu}}{\sqrt{g^2 +
\frac{4e^2}{3}}}$ field is massless in the CFL phase. In paper \cite{ABR}
the mixing angle obtained was twice times more $\alpha_{ABR} = \arccos \frac{g} {\sqrt{g^2 + e^2/3}}
\approx \frac{e}{\sqrt{3}g} \approx 1/20$ for the CFL phase. The discrepancy of our result with that of \cite{ABR}
is because the authors of \cite{ABR} used color generators $T^{a}$ normalized to
2, meanwhile, we used the standard definition of color generators $T^{a} = \frac{\lambda^a}{2}$
(of course, our result
can be easily recovered if we replace $g$ by $g$/2 in \cite{ABR}). According, e.g.,
to Particle Data Group \cite{PDG}, the standard definition of covariant derivative in QCD, which defines
strong coupling constant properly normalized at a fixed scale, is $D_{\mu} = \partial_{\mu} + igA_{\mu}^{a}T^{a}$,
where $T^a = \lambda^{a}/2$ and $\lambda^{a}$ are the Gell-Mann matrices, therefore,
the color generators $T^a$ are
normalized to $\frac{1}{2}$. Thus, we conclude that the correct value
of the mixing angle in the CFL phase is $\alpha_{CFL} = \arccos \frac{g} {\sqrt{g^2 + 4e^2/3}} \approx 1/10$. Although
our mixing angle is two times more than the one
found in \cite{ABR}, it does not change qualitative conclusions of \cite{ABR} because the mixing angle is still
a small number and
only a small part of the magnetic field flux is excluded inside the color superconducting region.

Let us now return to the question posed above about why the mixing angles in the CFL phase and the 2SC phase differ even
if the CFL gap goes to the 2SC gap for $m_s \to \infty$. Let us recall that the 2SC gap is
$\Delta_{\alpha \beta}^{ij}=\Delta \epsilon^{ij} \epsilon_{\alpha \beta 3}$, where flavor indices
run over 1 and 2. It is easy to check that the  operator $\tilde{Q} = Q -(T^3 + \frac{T^8}{\sqrt{3}})$, which
is equal to zero acting on the CFL gap, is also equal to zero acting on the 2SC gap.  Nevertheless, according to \cite{ABR}
the mixing angles in the CFL and 2SC phases differ. Why? The answer is that the generator $T^3$
is equal to zero acting on the 2SC gap in difference to the case of the CFL phase. Therefore, we can add this
generator to our $\tilde{Q}$ with any coefficient, i.e., the corresponding equation for $\tilde{Q}$ does not have a unique
solution. Obviously, the different choice of coefficients of color generators, which are equal to zero acting on the gap,
gives the different value of mixing angle. To find what is the correct $\tilde{Q}$ in this case,
we use the condition of minimum of energy. The less the mixing angle, the larger part of the external magnetic field
flux penetrates inside a color superconductor. The minimum of the mixing angle $\alpha=
\arccos\frac{g}{\sqrt{g^2 + \sum_{i=1}^8a_i^2e^2}}$ obviously corresponds to the minimum of $\sum_{i=1}^8a_i^2$.
For the 2SC phase, it is easy to show that the minimum of energy is given by
$\tilde{Q} = Q - \frac{T^8}{\sqrt{3}}$. (Note also that the choice of the diagonal SU(3) generators in the natural
representation of the group is ambiguous. If we consider the $T^3$ and $T^8$ used in \cite{ABR}, then we obtain
$\tilde{Q} = Q - \frac{1}{2}(T^3 + \frac{T^8}{\sqrt{3}})$, which obviously gives the same $\sum_{i=1}^8a_i^2$, i.e.,
the same mixing angle.) Thus, the mixing angle in the 2SC phase is indeed two times less
$\alpha_{2SC} \approx \frac{e}{\sqrt{3}g} \approx \frac{1}{20}$ than in the CFL phase.

\vspace{1cm}

\section{Explicit flavor symmetry breaking.}

\vspace{1cm}

Color and flavor symmetries are spontaneously broken in the CFL phase with $m_s=0$ or $m_s \ne 0$.
However, since color and flavor transformations are symmetries of the theory, all color and flavor transformed gaps
$U\Delta$ have the same energy
and, therefore, $SU_L(3) \times SU_R(3) \times SU_c(3) \times U_B(1) / SU_{L+R+c}(3)$ and
$SU_L(2) \times SU_R(2) \times SU_c(3) \times U_B(1) / SU_{L+R+c}(3)$ are the corresponding coset spaces of possible
solutions for gaps in the CFL phase with 3 massless quarks and in the CFL phase with $m_s \ne 0$, respectively.

Since the generator of electromagnetic transformations does not commute with SU(3) or SU(2) flavor transformations, 
one can expect that a color superconductor in the CFL phase chooses a specific value in its coset space of possible
solutions in external electromagnetic field. Indeed, if $A_{\mu} \ne 0$, then the Lagrangian has only
$SU_c(3)$ color symmetry. Thus, the flavor symmetry is explicitly broken if $A_{\mu} \ne 0$.
We show below that since the CFL gap
locks flavor and color, one can, in fact, use color transformations and, therefore, the corresponding flavor transformed
gap $U\Delta$ has the same energy as the initial gap $\Delta$ even if $A_{\mu} \ne 0$
(the case of a color transformed gap is, of course, trivial because Q commutes with color transformations).

We first consider the CFL phase with 3 massless quarks. The corresponding gap is \cite{activity}
\begin{equation}
\Delta_{\alpha \beta}^{ij} = k_1 \delta^i_{\alpha}\delta^j_{\beta} + k_2 \delta^i_{\beta}\delta^j_{\alpha}.
\end{equation}
The operator $\tilde{Q} = Q - (T^3 + \frac{T^8}{\sqrt{3}})$ is equal to zero acting on the gap. Let us consider a flavor
transformed gap $U\Delta$. Since $[Q,U] \ne 0$, $\tilde{Q}U\Delta \ne 0$ in the general case and we should seek
another operator
$\tilde{Q_U}$, which is equal to zero acting on the gap $U\Delta$, i.e., we seek a solution of
the equation
\begin{equation}
(Q_{t.r.} + a_1T^1_{t.r.} + ... +  a_8T^8_{t.r.})U_{t.r.}\Delta = 0,
\end{equation}
which gives
\begin{equation}
Q + \sum_{i=1}^8 a_iU(T^i)^TU^+ = 0
\end{equation}
or what is more convenient for analysis
\begin{equation}
U^+QU + \sum_{i=1}^8 a_i(T^i)^T = 0,
\end{equation}
where we used the fact that $k_1$ and $k_2$ are independent order parameters.
Multipling Eq.(6) by $T^j$ and taking trace, we obtain a system of equations for $a_i$
\begin{equation}
\sum_{i=1}^8 a_i tr((T^i)^TT^j) = - tr(U^+QUT^j).
\end{equation}
It is easy to check that for U=1 we obtain the old solution $a_3 = -1, a_8 = -\frac{1}{\sqrt{3}}$, and others $a_i=0$.
Our analysis is simplified by noting that Q can be represented as $Q = -\frac{I}{3} + A$, where $I$ is the unity
matrix and $A$ is a matrix whose the only nonzero element is $A_{11}=1$. 
Indeed, since $U^+U = 1$ and $trT^i = 0$, we need to calculate only $trU^+AUT^j$ on the right-hand side of Eq.(7).
By parametrizing U as follows
\[
\begin{array}{rrr}
 & & \\
 &U&= \\
 & & \\
 \end{array}
\left(
\begin{array}{rrr}
u_1 & u_2 & u_3 \\
v_1 & v_2 & v_3 \\
w_1 & w_2 & w_3 \\
\end{array} \right),
\]
we find $a_i$ and then $\sum_{i=1}^8a_i^2$, which is equal to
\begin{equation}
\sum_{i=1}^8a_i^2 = \frac{4}{3}(|u_1|^2 + |v_1|^2 + |w_1|^2)^2.
\end{equation} 
Since U is a unitary matrix, we have $UU^+ = 1$ that gives us $|u_1|^2 + |v_1|^2 + |w_1|^2 = 1$.
Therefore, we obtain from Eq.(8) that $\sum_{i=1}^8a_i^2=4/3$ for any
flavor transformed gap. Obviously that a similar analysis can be used for
the CFL phase with $m_s \ne 0$. The case of the 2SC phase is trivial because flavor symmetry is not spontaneously broken
in this phase. Thus, although the generator of electromagnetic transformations does not commute with flavor
transformations and we have an explicit flavor symmetry breaking, all flavor transformed CFL gaps have the same energy.
Of course, the reason for this is that the CFL gap locks flavor and color and, therefore, the corresponding equation
for the operator $\tilde{Q_U}$, which is equal to zero acting on $U\Delta$, coincides with the equation for
the operator $\tilde{Q}$ (with unitary transformed color generators $U^+T^iU$ (see Eq.(5))), which is equal to zero
acting on the $\Delta$, that, of course, gives us the same $\sum_{i=1}^8a_i^2$ because color generators are
defined up to a unitary transformation.

\vspace{1cm}

\section{Conclusions.}

\vspace{1cm}

We considered how external magnetic field influences color superconductivity for the CFL phase with 3
massless quarks, the CFL phase with $m_s \ne 0$, and the 2SC phase with 2 massless quarks. We explained why
the mixing angles in the 2SC and CFL phases are different even if the gap of the CFL phase with $m_s \ne 0$ goes
to the 2SC gap for $m_s \to \infty$. We
showed that despite expilicit flavor symmetry breaking in external magnetic field, all values of
flavor transformed gaps in their coset spaces of possible solutions in the CFL phase are equivalent. 

The author thanks A.A. Natale for helpful discussions and encouragement. I am grateful to
V.P. Gusynin and V.A. Miransky for reading the manuscript and critical remarks.
The author thanks K. Rajagopal for comments on the manuscript.
I acknowledge useful conversations with M. Nowakowski and I. Vancea.
This work was supported in part by FAPESP grant No. 98/06452-9.

\end{document}